\documentclass{emulateapj}

\usepackage{natbib} 
\begin{document}
\def\mpch {$h^{-1}$ Mpc} 
\def\kpch {$h^{-1}$ kpc} 
\def\kms {km s$^{-1}$} 
\def\lcdm {$\Lambda$CDM } 
\def\xir {$\xi(r)$}
\def\wprp {$w_p(r_p)$}
\def\xisp {$\xi(r_p,\pi)$}
\def\xis {$\xi(s)$}
\def\rr {$r_0$}
\def\etal {et al.}

\slugcomment{Accepted to ApJ}

\title{The DEEP2 Galaxy Redshift Survey: Clustering of Quasars and 
Galaxies at $z=1$}
\author{Alison L. Coil\altaffilmark{1,2,3}, 
Joseph F. Hennawi,\altaffilmark{1,2}
Jeffrey A. Newman\altaffilmark{1,4},
Michael C. Cooper\altaffilmark{2}, 
Marc Davis\altaffilmark{2},
}
\altaffiltext{1}{Hubble Fellow}
\altaffiltext{2}{Department of Astronomy, University of California,
Berkeley, CA 94720 -- 3411} 
\altaffiltext{3}{Steward Observatory, University of Arizona,
Tucson, AZ 85721}
\altaffiltext{4}{Institute for Nuclear and Particle Astrophysics,
  Lawrence Berkeley National Laboratory, Berkeley, CA 94720}

\begin{abstract}

We present the clustering of DEEP2 galaxies at $0.7<z<1.4$ around
quasars identified using both the SDSS and DEEP2 surveys.  We
measure the two-point cross-correlation of a sample of 36
optically-selected, spectroscopically-identified quasars from the
SDSS and 16 more found in the DEEP2 survey with the full DEEP2
galaxy sample over scales $0.1 < r_p < 10$ \mpch.  The clustering
amplitude is found to be similar to the auto-correlation function of
DEEP2 galaxies, with a relative bias of $b=0.89 \pm0.24$ between
quasars and DEEP2 galaxies at $z\sim1$.  No significant dependence
is found on scale, quasar luminosity, or redshift over the ranges we
probe here.  The clustering amplitude errors are comparable to those
from significantly larger quasar samples, such as the 2dF QSO
Redshift Survey.  This results from the statistical power of
cross-correlation techniques, which exploit the fact that galaxies
are much more numerous than quasars.  We also measure the local
environments of quasars using the $3^{\rm rd}$-nearest-neighbor
surface density of surrounding DEEP2 galaxies.  Quasars are found in
regions of similar mean overdensity as blue DEEP2 galaxies; they
differ in environment from the red DEEP2 galaxy population at
$2\sigma$ significance.  Our results imply that quasars do not
reside in particularly massive dark matter halos at these redshifts,
with a mean dark matter halo mass of $M_{200}\sim3 \times10^{12}
M_\sun$ in a concordance \lcdm cosmology.
  
\end{abstract}


\section{Introduction}

There is growing evidence that most galaxies have a supermassive black
hole in their nucleus \citep[for a review, see e.g.][]{Richstone98}.
Accretion onto these supermassive black holes likely powers quasars,
which now appear to be highly relevant for galaxy formation and
evolution models.  In particular, the observed correlation between
black hole mass and the velocity dispersion of stars in the bulge
components of galaxies \citep{Gebhardt00,Ferrarese00} indicates some
form of feedback or connection between the growth of black holes and
their parent galaxies.  In addition, quasar or active galactic nuclei 
(AGN) feedback may play a significant role in reproducing the
observed color-magnitude diagram at high redshift \citep{Springel05,Croton06}.

It remains unclear what physical mechanism fuels quasars and
what impact their presence has on galaxy formation and evolution.
Quasar clustering measurements can be used to infer their lifetimes and
large-scale environments, which help address these issues.  Clustering
measurements lead to an estimate of the mean dark matter halo mass of
a given population, by using semi-analytic methods or 
cosmological dark matter N-body
simulations to relate the observed number density and clustering
strength of a data sample to its parent dark matter halo population
\citep[e.g.,][]{Kaiser84, Efstathiou88, Cole89, Mo96, Sheth99}.  
This allows the population to be
placed in a cosmological context, useful for comparisons with
simulations and galaxy evolution models. Quasar clustering analyses 
can further constrain the physics behind the creation and
fueling of quasars from the lifetimes inferred by the ratio of the 
quasar number densities to those of their parent dark matter halos
\citep{Haiman01,Martini01,Wyithe05}.

The advent of the 2dF QSO Redshift Survey \citep{Croom04a} allowed
the first robust measurements of the quasar clustering amplitude using
a large sample covering sizeable areas of the sky, $\sim20,000$
objects in $\sim700$ square degrees \citep{Croom02,Porciani04,Croom05}.  
Their main results are: 1)
quasars between redshift $0.3<z<2.2$ show similar clustering as local
galaxies, well-fit by a power-law with a clustering scale-length of
$r_0=4.8 \ (+0.9/-1.5)$ \mpch \ and a slope of $\gamma=1.5 \pm0.2$; 2) 
the clustering amplitude shows little dependence on quasar luminosity
(within the large error bars); and 3) that amplitude 
increases with redshift from $0.5 < \bar{z} < 2.5$.
\citet{Myers06} confirm these conclusions using the projected angular
clustering of $\sim$80,000 quasars in the Sloan Digital Sky Survey (SDSS).
These results generally support a picture in which quasar luminosity is not 
strongly correlated with the parent dark matter halo mass, and 
characteristic 
quasar host halo masses are $\sim10^{12}-10^{13} M_\sun$ at $z\sim1-2$.
Additionally, only a few percent of the potential parent halos 
actually host a quasar \citep{Porciani04}, and the inferred 
quasar lifetime is not long, $\sim10^7$ years.

Unfortunately, the error bars on the clustering amplitudes estimated
from the quasar auto-correlation function are still large, due to their 
relatively low number density.  Studying the clustering of
galaxies around quasars (by measuring the cross-correlation function
of quasars and galaxies) can afford more precise measurements of the
clustering amplitude of quasars, as the number density of galaxies is
much higher \citep[e.g.,][]{Kauffmann02}.  It also provides a
measure of the local environment in which quasars reside, which is
relevant to understanding the physics of quasar fueling.

It has long been known that quasars are associated with enhancements
in the distribution of galaxies
\citep{BSG69,Seldner79,YG84,YG87,BC91,LS95,SBM95,HG98,SBM00}. However,
there has been significant scatter in these measurements of
quasar-galaxy correlations \citep[see][Table 1 for a compilation of
recent studies]{BBW01} caused by heterogeneous quasar samples,
methodology, and imaging depths.  Large surveys such as the 2dF
and SDSS provide homogeneous samples of $\sim 10,000$ quasars/AGN 
surrounded by $\sim 100,000$ galaxies, providing an opportunity to 
robustly study the clustering of galaxies around low redshift quasars.  

For quasars at $z< 0.3$, \citet{Croom04} show that the quasar-galaxy
cross-correlation function in 2dF data is the same as the galaxy
auto-correlation function, measured in redshift space 
on scales of $s=1-20$ \mpch.  \cite{Wake04}
measure the clustering amplitude of AGN in redshift space for $z<0.2$
in the SDSS DR1 data, where AGN are selected using classic
emission-line ratio diagnostics and they have excluded broad-line
quasars.  They find that AGN have a similar clustering strength as
galaxies in the same redshift range, with a small anti-bias with
respect to the full SDSS galaxy sample of $b=0.92 \pm0.03$ on scales
$s\sim0.4-20$ \mpch.  Comparing their results to dark matter
simulations, they conclude that the minimum host dark matter halo mass
for these AGN is $M_{min}=2 \times 10^{12} M_\sun$. \cite{Constantin06} 
compare the
redshift-space auto-correlation function of narrow-line AGN in SDSS
data and find that Seyferts are less clustered than the full galaxy
sample, while LINERS show similar clustering properties to all
galaxies. \citep{Serber06} study the environments of $z<0.4$
luminous ($M_{i}<-22$) quasars in the SDSS and find that quasars
cluster similarly to $L^*$ galaxies on scales $\sim1$ \mpch, in
agreement with the aforementioned studies.  However, on smaller scales
at $r<100$ \kpch \ they find that quasars reside in regions overdense by
a factor of $1.4-3$ compared to regions around $L^*$ galaxies, with the larger
density enhancement occurring for the most luminous quasars
($M_i<-23.3$) in their sample.

At higher redshifts \cite{Adelberger05a} measure the cross-correlation
of 79 AGN and $\sim$1600 Lyman-break galaxies at $2<z<3$.  They find
that the cross-correlation scale-length is $r_0\sim5 \pm2$ \mpch,
similar to the clustering amplitude of Lyman-break galaxies
themselves, and does not depend on AGN luminosity, from which they
infer that brighter and fainter AGN reside in dark matter halos of
similar mass and therefore fainter AGN are longer lived.

In this paper we present the first measurements of the quasar-galaxy
cross-correlation function at $z\sim1$, using data from the SDSS and
DEEP2 galaxy redshift surveys. Because the selection function of DEEP2
galaxies has been precisely quantified, we are able to robustly
correct for small-scale selection biases in the galaxy sample allowing
us to measure the clustering strength as a function of scale from
$0.1<r<10$ \mpch. 
We also present measurements of the mean local environment of quasars
at $z\sim1$, compared to galaxies in the DEEP2 survey. The layout of
the paper is as follows: \S 2 presents the quasar and galaxy data
samples; \S 3 discusses the methods used to estimate the
cross-correlation function; clustering and environment results are
presented in \S 4 and \S 5 and discussed in \S 6.

\section{SDSS Quasar and DEEP2 Galaxy Samples}

We use data from both the SDSS and DEEP2 redshift surveys where they
overlap -- ie., in the DEEP2 fields -- in the redshift range
$0.7<z<1.4$.  There are 36 spectroscopically-identified quasars from
the SDSS and an additional 16 quasars in the DEEP2 Galaxy Redshift
Survey  in the 
volume sampled by DEEP2 galaxies.  We measure
cross-correlation statistics for three quasar/AGN samples: 1) all SDSS
quasars, 2) all SDSS and DEEP2 quasars brighter than $M_B<-22$, and 3)
all SDSS and DEEP2 quasars and an additional 7 DEEP2 type 1
AGN (labeled as the 'all quasars' sample).

Redshifts and $M_B$ magnitudes for quasars in both the SDSS and
DEEP2 samples are shown in Fig. \ref{QSOprops}, and details of each
sample are given below.  The median redshifts of the SDSS quasar
sample,  the DEEP2 quasar and AGN sample, and the full combined sample 
are  $z_{median}=1.01$, 0.80, and 0.99, respectively.  
The redshifts of the quasars used in this study were computed via
cross-correlation with a composite quasar template. For the SDSS
quasars, the spectral coverage is 3800-10000 \AA; in the redshift
range $z=0.7-1.4$ the MgII 2798 \AA \ and [OIII] 5007 \AA \ emission lines
dominate measurements of the quasar redshift.  For
the DEEP2 observations, the smaller spectral coverage of
$\sim$6200-9400 \AA \
results in either MgII or [OIII] being present.
\cite{Boroson05} find that the [OIII] emission line has
an average shift of 40 \kms \ and a dispersion of $<$100 \kms \ 
about the systemic reference frame defined by low-ionization forbidden lines.
\cite{Richards02b} find that the MgII emission line has a median shift of 97
\kms \ and a dispersion of 269 \kms \ about [OIII] (assumed to be 
systemic). Thus a very 
conservative estimate of the errors in the quasar redshifts due to
both shifts from systemic and errors in the redshift determination
would be $<$ 500 \kms, or $dz < 0.003$, which corresponds to 5 \mpch \ 
at $z = 1.0$.

\subsection{The SDSS Quasar Sample}

The SDSS uses a dedicated 2.5m telescope and a
large format CCD camera \citep{Gunn98,Gunn06} at the Apache Point Observatory
in New Mexico to obtain images in five broad bands \citep[$u$, $g$,
$r$, $i$ and $z$, centered at 3551, 4686, 6166, 7480 and 8932 \AA,
respectively;][]{Fukugita96,Stoughton02} of the high Galactic latitude sky in
the Northern Galactic Cap. Based on these imaging data, spectroscopic
targets chosen by various selection algorithms are observed with two double
spectrographs producing spectra covering \hbox{3800--9200 \AA} 
with a spectral resolution ranging from 1800 to 2100.  Details of the
spectroscopic observations can be found in \citet{York00},
\citet{Castander01}, and \citet{Stoughton02}.  Additional details on
the SDSS data products can be found in 
\citet{Abazajian03,Abazajian04,Abazajian05b}.

We briefly summarize the essential details about the SDSS quasar
catalog, and refer the reader to \citep{Schneider05} for a more
thorough discussion.  In the redshift range of interest to us here
$0.7<z<1.4$, the primary SDSS low redshift target selection algorithm
\citet{Richards02} imposes an $i$ magnitude limit of~19.1, and the
SDSS quasar catalog has very high completeness \citet{vanden05} above
this limit.  Supplementing this primary quasar selection are quasars
targeted by other SDSS target selection criteria \citep{Blanton03} and
most of the quasars with $0.7<z<1.4$ that have $i>19.1$ were selected
in this way, although no attempt at completeness is made for these
serendipitous targets.  The official SDSS Third Data Release Quasar
Catalog contains 46,420 quasars \citep{Schneider05}.  We have used an
unofficial quasar catalog based on the Princeton/MIT spectroscopic
reductions \footnote{Available at http://spectro.princeton.edu}
\citep{Schlegel06}, which differs slightly from the official catalog
\footnote{Spectra in the Princeton/MIT reductions are cross-correlated
with several spectral templates (quasar, star, various galaxy types).
A quasar is defined to be any object whose $\chi^2$ difference with
the quasar template is below a specified value.}.

The quasar catalog was matched to the DEEP2 survey area, resulting in
a total of 36 quasars in the redshift range $0.7<z< 1.4$.  Six of the
quasars were also observed by the DEEP2 Galaxy Redshift Survey.
Spectra of all of these objects were visually inspected to verify that
they were indeed broad-line quasars at the specified redshift.
Coordinates, redshift, absolute magnitudes, and SDSS photometry are
presented in Table~\ref{table:qsos}.  Absolute $M_B$ magnitudes are
computed from the cross-filter K-correction, $K_{Bi}(z)$ between SDSS
i-band and Johnson B-band using the composite quasar spectrum of
\citet{Vanden01} Johnson-B and SDSS $i$ filter curves.  If we restrict
the sample to only those quasars above the SDSS flux $i<19.1$ (for low
redshift quasars), we would be left with only 12 objects in the DEEP2
area, making a statistically significant measurement of quasar-galaxy
clustering very difficult.  To maximize the number of quasars, we thus
considered all 36; however, the incompleteness of the SDSS at these
fainter magnitudes implies that we do not know their redshift or
angular selection functions.  In \S~\ref{sec:cross} we describe how we
overcome this unknown quasar selection function to compute the
quasar-galaxy cross-correlation function.

\subsection{The DEEP2 Quasar and Galaxy Samples}

The DEEP2 Galaxy Redshift Survey is a three-year project using the
DEIMOS spectrograph \citep{Faber03} on the 10m Keck II telescope to
survey optical galaxies at $z\simeq1$ in a comoving volume of
approximately 5$\times$10$^6$ $h^{-3}$ Mpc$^3$.  Using $\sim1$~hr
exposure times, the survey has measured redshifts for $\sim30,000$
galaxies in the redshift range $0.7<z<1.5$ to a limiting
magnitude of $R_{\rm AB}=24.1$ \citep{Coil03xisp, Faber06}.  The
survey covers three square degrees of the sky over four widely
separated fields to limit the impact of cosmic variance.  Due to the
high resolution ($R\sim5,000$) of the DEEP2 spectra, redshift errors,
determined from repeated observations, are $\sim30$km s$^{-1}$.
Details of the DEEP2 observations, catalog construction and data
reduction can be found in \citet{Davis03, Coil04, Davis04, Faber06}.
Restframe $(U-B)$ colors have been derived as described in
\citep{Willmer06}; here they are in AB magnitudes.

In the DEEP2 dataset we have spectroscopically identified 9 additional
broad-line quasars which were not observed by SDSS and another 7 type
1 AGN with both broad and narrow emission lines in the redshift range
$0.7<z<1.4$; their properties are listed in Table 2. For the rest of
this paper we refer to these objects solely as quasars.  For objects
with SDSS $i<21.0$, absolute $M_B$ magnitudes were computed as above for
the SDSS quasars; others use the observed DEEP2 R magnitude to
estimate K-corrections.  Due to the limited spectral range observed by
DEEP2, quasars between $0.9<z<1.3$ are not likely to be identified as
no broad emission lines would be visible in the spectra.  We use as a
galaxy sample for cross-correlation purposes all DEEP2 galaxies within
30 \mpch \ of the SDSS and DEEP2 quasars; this results in a sample of
$\sim$5000 DEEP2 galaxies.  Because the clustering of the full DEEP2
flux-limited galaxy sample happens to be flat with redshift between
$0.7<z<1.4$, there are no strong biases introduced as a function of
redshift by using the full galaxy sample.

To convert measured redshifts to comoving distances along the line of
sight we assume a flat \lcdm cosmology with $\Omega_{\rm m}=0.3$ and
$\Omega_{\Lambda}=0.7$.  We define $h \equiv {\rm {\it H}_0/(100 \ km
\ s^{-1} \ Mpc^{-1}})$ and quote correlation lengths, \rr, in comoving
\mpch.

\section{Measuring the Cross-Correlation Function}
\label{sec:cross}

The two-point auto-correlation function \xir \ is defined as a measure
of the excess probability above Poisson of finding an object in a
volume element $dV$ at a separation $r$ from another randomly chosen
object,
\begin{equation}
dP = n [1+\xi(r)] dV,
\end{equation}
where $n$ is the mean number density of the object in question
\citep{Peebles80}.  The cross-correlation function is the excess
probability above Poisson of finding an object from a given sample at
a separation $r$ from a random object in another sample.  Here we
measure the cross-correlation between quasars and galaxies:
\begin{equation}
dP(G|Q) = n_G [1+\xi_{QG}(r)] dV,
\end{equation}
which is the probability of finding a galaxy ($G$) in a volume element
$dV$ at a separation $r$ from a quasar ($Q$), where $n_G$ is the
number density of galaxies.

To estimate the cross-correlation function between our quasar and
galaxy samples, we measure the observed number of galaxies around each
quasar as a function of distance, divided by the expected number of
galaxies for a random distribution.  We use the estimator 
\begin{equation}
\xi=\frac{QG}{QR}-1,
\end{equation}
where $QG$ are quasar-galaxy pairs and $QR$ are quasar-random pairs at
a given separation, where the pair counts have been normalized by
$n_G$ and $n_R$, respectively, the mean number densities in the full galaxy
and random catalogs.  This estimator is preferred here as it does not
require knowledge of the quasar selection function, only the galaxy
selection function, which is well-quantified.  
For each quasar we create a random catalog with
the same redshift distribution as all DEEP2 galaxies and the same 
sky coverage as the DEEP2 galaxies in that field, applying
the two-dimensional window function of the DEEP2 data in the plane of
the sky.  Our redshift success rate is not entirely uniform across the
survey; some slitmasks are observed under better conditions than
others and therefore yield a higher completeness.  This
spatially-varying redshift success rate is taken into account in the
spatial window function.  We also mask the regions of the random
catalog where the photometric data are affected by saturated stars or CCD
defects.

Redshift-space distortions due to peculiar velocities along the
line-of-sight and uncertainties in the systemic redshifts of the
  quasars \citep{Richards02} will introduce systematic effects to the
estimate of \xir.  To uncover the real-space clustering properties of
galaxies around quasars we measure $\xi$ in two dimensions, both
perpendicular to ($r_p$) and along ($\pi$) the line of sight. 
In applying the above estimator to galaxies, pair counts are
computed over a two-dimensional grid of separations to estimate \xisp.
To recover \xir, \xisp \ is integrated along the $\pi$ direction and 
projected along the $r_p$ axis.  As
redshift-space distortions affect only the line-of-sight component of
\xisp, integrating over the $\pi$ direction leads to a statistic
\wprp, which is independent of redshift-space distortions.  Following
\cite{Davis83},
\begin{equation}
w_p(r_p)=2 \int_{0}^{\infty} d\pi \ \xi(r_p,\pi)=2 \int_{0}^{\infty}
dy \ \xi(r_p^2+y^2)^{1/2},
\label{eqn}
\end{equation}
where $y$ is the real-space separation along the line of sight. 
Here we integrate to a maximum separation in the $\pi$ 
direction of 20 \mpch, as the signal to noise degrades quickly for larger 
separations where $\xi$ becomes small.  

If \xir \ is modeled as a power-law, $\xi(r)=(r/r_0)^{-\gamma}$, and
\wprp \ is integrated to $\pi_{max}=\infty$, then \rr
\ and $\gamma$ can be extracted from the projected correlation
function, \wprp, using an analytic solution to Equation \ref{eqn}:
\begin{equation}
w_p(r_p)=r_p \left(\frac{r_0}{r_p}\right)^\gamma
\frac{\Gamma(\frac{1}{2})\Gamma(\frac{\gamma-1}{2})}{\Gamma(\frac{\gamma}{2})},
\label{powerlawwprp}
\end{equation}
where $\Gamma$ is the gamma function.  A power-law fit to \wprp \ will
then recover \rr \ and $\gamma$ for the real-space correlation
function, \xir.  However, \xir \ is not expected to be a power-law to
very large scales, and we have integrated to $\pi_{max}=20$ \mpch, not $\infty$.
Instead, we recover \rr \ and $\gamma$ by numerically integrating
Eqn. 4 to $\pi_{max}=20$ \mpch \ and determining the values of \rr \
and $\gamma$ which minimize $\chi^2$.  We include redshift-space
distortions on large scales due to coherent infall of galaxies
(\cite{Kaiser87}, see \cite{Hamilton92} and section 4.1 
of \cite{Hawkins03} for the relevant equations for the correlation
function), where for $\beta=\Omega_m^{0.6}/b$ we assume a linear bias 
relative to the dark matter of $b=1.3$ (see \cite{Coil06lum} for the bias of
DEEP2 galaxies) and $\Omega_m=0.24$ at $z=0$ \citep{Spergel06}.  
This method assumes that \xir \ is a power-law only to a scale of 
$\pi_{max}$ and results in \rr \ and $\gamma$ values within a few \% 
of those obtained using Eqn. 5, for our value of $\pi_{max}=20$ \mpch.  
Deviations from Eqn. 5 are significant only on larger scales, where
$r_p/\pi_{max} \gtrsim 0.25$. 
We note that comparisons to simulations or models 
that directly compute \wprp \ to the same $\pi_{max}$ as in
the data, such as that of \cite{Conroy06}, do not suffer from this
effect as they do not use the quoted power-law fits to the data. 

To test directly the possible effects of quasar redshift errors, we
convolve $\xi(r_p,\pi)$ with a 500 \kms \ Gaussian (the upper limit of
the actual error) in the $\pi$ 
direction upon the integral from $-\pi_{max}$ to $+\pi_{max}$.
For $\pi_{max}=20$ \mpch \ we find that the measured \wprp \ is 0.4\% lower
at $r_p=1$ \mpch \ and 2\% lower at $r_p=10$ \mpch \ than if there
were no redshift errors.  We therefore conclude that redshift errors
are negligable.

\section{Quasar-Galaxy Clustering Results}

We show the cross-correlation function results in Fig. \ref{wprp}.
The left panel is the projected cross-correlation function between the
SDSS quasar sample and DEEP2 galaxies; the dashed line is the observed
correlation function and the solid line has been corrected for
slitmask target effects using the mock catalogs of \cite{Yan03}. For a
full discussion of the slitmask targeting effect see Section 3.3 of
\cite{Coil03xisp}. Briefly, the issue is that when observing galaxies
with multi-object slitmasks, the spectra cannot overlap on the CCD
array; therefore, objects that lie near each other in the direction on
the sky that maps to the wavelength direction on the CCD cannot be
simultaneously observed.  This will necessarily result in
under-sampling the regions with the highest density of targets on the
plane of the sky, which leads to underestimating the correlation
function on small scales.  The effect is not large (as can be seen in
the left panel of Fig. \ref{wprp}) and for the DEEP2 survey leads to
underestimating the correlation length by 2-3\% \citep{Coil06lum}.  To
correct for this effect we use the ratio of the projected
cross-correlation function in the mock catalogs between a sample of
randomly-selected galaxies in the catalog (acting as a quasar sample)
with the full sample of other galaxies and a subsample that would have
been selected to be observed on slitmasks after the slitmask targeting
code is applied.  For the SDSS quasar sample, the correction is
smaller than for the DEEP2 quasar sample, as the SDSS objects do not suffer
slit collisions with DEEP2 targets.  We therefore correct for the SDSS
sample using random galaxies in the mock catalogs before target
selection (acting as SDSS quasars) with galaxies after target selection
(acting as DEEP2 galaxies), while for the DEEP2 quasars we use
galaxies in the mock catalogs after target selection for both the 
quasar and tracer samples.

Error bars on \wprp \ are estimated in two ways; the solid error bars
are derived from jacknife resampling the quasar sample 
while the dotted error bars show the
standard deviation across twelve independent mock catalogs.  To derive
the error in the mock catalogs, we use 45 randomly-selected galaxies
in each catalog (chosen before the slitmask target selection is
applied) to act as proxies for quasars.  We then calculate the 
cross-correlation
function with the rest of the galaxy sample, in the same manner as is
calculated for the DEEP2 data.  The error in the mock catalogs are
quite comparable to the jacknife errors; we use the jacknife errors
throughout this paper to be conservative, as they are slightly larger.

The right panel of Fig. \ref{wprp} shows results for each of the three
quasar samples, where the errors shown are from jacknife resampling
within each dataset.  There are no statistically significant
differences in the bias in the three samples, which reflects a lack of
luminosity-dependence in the quasar-galaxy cross-correlation for the
range of quasar luminosities that we probe here.  We also split the SDSS quasar
sample by luminosity at $M_B=-23.5$ and find no difference at the 1 $\sigma$ level 
between the fainter and brighter quasar samples, which have median magnitudes of
$M_B=-22.8$ and $M_B=-24.1$.
\citet{Croom05} similarly fail to find any dependence of the quasar
 auto-correlation on luminosity.

Power-law fits are derived for each sample and results are given in
Table \ref{restable}.  Fits are given over the full $r_p$ range
($0.1<r_p<10$ \mpch) and for the larger scales only, $1<r_p<10$ \mpch.
Differences in fits on the two different scale ranges result in $r_0$
and $\gamma$ values that are within 1 $\sigma$ of each other;
fitting just on larger scales results in a somewhat steeper slope and
larger correlation length.

We calculate the relative bias of quasars to DEEP2 galaxies by
dividing the quasar-galaxy cross-correlation function by the
auto-correlation function of DEEP2 galaxies, shown in the left panel
of Fig \ref{bias}.  To minimize differences in the galaxy population
used for the cross-correlation function with quasars and the galaxy
auto-correlation function, we calculate \wprp \ for the DEEP2 galaxies
here by selecting 30 random galaxies around each quasar (1560 galaxies
in total), within a half-length of $\Delta z=0.1$ and in the same
field in the plane of the sky, and then measuring the
cross-correlation of those galaxies with all of the surrounding DEEP2
galaxies.  This ensures that the galaxies used in both the
quasar-galaxy cross-correlation and the galaxy auto-correlation
functions have the same redshift, magnitude and color distributions,
and further ensures that the same volume has been used in both
measurements, reducing cosmic variance errors in the relative bias.
However, using the auto-correlation function of the full DEEP2 galaxy
sample over all redshifts does not change the results.  The right
panel of Fig. \ref{bias} shows the relative bias of quasars to DEEP2
galaxies as a function of scale.  The mean relative bias on scales
$0.1<r_p<10$ \mpch \ and $1<r_p<10$ \mpch \ is given in Table
\ref{restable}; there is no significant scale-dependence.  The errors
on the relative bias are estimated from jacknife resampling of bias
measurements, which include covariance between adjacent $r_p$ bins.
The relative bias is found to be $\sim0.90 \pm0.25$ for each of the
samples, consistent with the quasar samples having the same clustering
amplitude as DEEP2 galaxies.  We have further divided the quasar
sample into two redshift bins, $0.7<z<1.0$ and $1.0<z<1.4$, and find
no difference in the bias between the two samples.

We further show the relative bias of red and blue DEEP2 galaxies to
the DEEP2 galaxy sample in Fig. \ref{bias}, where we have again
measured the cross-correlation of the same 30 random galaxies within
$\Delta z=0.1$ of each quasar (1560 galaxies total) with the
surrounding DEEP2 galaxies with either red or blue colors, where the
division is defined as $(U-B)=1.0$, near the valley between the red
and blue populations of the DEEP2 color-magnitude diagram (see Fig. 3
of \cite{Cooper06}).  The relative biases for red and blue galaxies
are measured in the same way as for quasars, where we divide by the
auto-correlation function of DEEP2 galaxies.  The red galaxy relative
bias is $1.36 \pm0.10$ on scales $0.1<r_p<10$ \mpch \ and $1.41
\pm0.12$ on scales $1<r_p<10$ \mpch, $2\sigma$ higher than the quasar
relative bias, while the blue relative bias is $0.95 \pm0.02$ on the
same scales, consistent with the quasar relative bias.

We note that cleaner measures of the red and blue galaxy relative bias
are possible using the auto-correlation function of the full DEEP2
dataset over all redshifts, and the results on scales $r_p>1$ \mpch \ 
are consistent with what is found using the smaller galaxy sample
around quasars here. However, red galaxies in the DEEP2 data 
are seen to have a steeper
correlation slope than blue galaxies \nocite{Coil03xisp} (Coil et
al. 2004, Coil et al. in prep.), which is not reflected in this sample
(see the right panel of Fig. \ref{bias}).  The relative bias on scales
$r_p=1-10$ \mpch \ is consistent with what is seen here.

\section{Quasar Environments at $z\sim1$}

For each SDSS and DEEP2 quasar in our sample, we also estimate the
local environment using the $3^{\rm rd}$-nearest-neighbor surface
density of surrounding DEEP2 galaxies.  This estimator proved to be
the most robust indicator of local overdensity for high redshift surveys in the tests of
\cite{Cooper05}.  Like the projected cross-correlation function, this
statistic provides a measure of the density of galaxies surrounding
sources of a given type, such as quasars; however, it is measured on
an adaptive scale (typically $\sim2$ \mpch for DEEP2 samples), rather than as a function
of scale.

We measure the $3^{\rm rd}$-nearest-neighbor surface density of DEEP2 galaxies, 
$\Sigma_3$, within a line-of-sight velocity window
of $\pm1000\ {\rm km}/{\rm s}$;  it is related to
the projected distance to the $3^{\rm rd}$-nearest
neighbor, $D_{p,3}$, as $\Sigma_3=3 / (\pi D_{p,3}^2)$. We likewise 
measure the local surface density about individual DEEP2 galaxies for 
comparison samples; here we compute the mean overdensity for all DEEP2 galaxies and
blue and red galaxies (defined as in \S 3) separately. 
To correct for the redshift dependence of the sampling rate of the DEEP2 
survey, each surface density value is divided by the median $\Sigma_3$ of 
galaxies at that redshift; correcting the measured surface densities in this 
manner converts the $\Sigma_3$ value into measures of overdensity
relative to the median density (given by the notation $1+\delta_3$ here). 
For complete details on the determination of the galaxy environments and 
corrections for variations in selection in redshift and on the sky, 
we refer the reader to \cite{Cooper06}.  

In Table 4, we compare the mean overdensity of the quasar population 
to three DEEP2 galaxy samples, all of which are measured
over the redshift range  $0.75 < z < 1.35$. Errors are computed using the standard
deviation of the overdensity distribution divided by the square root of the 
number of objects.  The DEEP2 galaxy samples all have standard deviation $\sigma(log(1+\delta))=0.64$,
while the quasar sample has $\sigma(log(1+\delta))=0.57$. To be conservative,
we assume that the $\sigma$ on the quasar sample is low by chance and use the same
value as the galaxy samples.  We find that the mean  
local environment of the quasars is consistent with the mean 
environment of the full DEEP2 galaxy population and the blue galaxy 
population, while red galaxies in DEEP2 are found in more 
overdense environments than the quasars at a $2\sigma$ level.
This is quite consistent with the cross-correlation results from the previous section.

\section{Discussion and Conclusions}

We show that quasar host galaxies at $z\sim1$ have similar clustering
properties and local environments to typical DEEP2 galaxies.  This
implies that quasar host galaxies at $z\sim 1$ have dark matter halo
masses comparable to those of DEEP2 galaxies and are not strongly
biased relative to the galaxy population.  They have similar local
environments to and cluster much like blue, star-forming galaxies
rather than red galaxies.

We find that $r_0\sim3.4 \pm0.7$ \mpch \ for the quasar-galaxy
cross-correlation, while we have measured $r_0\sim3.75$ \mpch \ for
the full DEEP2 galaxy sample \citep{Coil06lum}.  Assuming a linear
bias, such that $\xi_{QG}=(\xi_{QQ}*\xi_{GG})^{0.5}$, then if $\gamma$
is identical our inferred quasar clustering scale-length is
$r_0\sim3.1 \pm0.6$ \mpch.  In \cite{Coil06lum} the DEEP2 galaxies as a
whole are shown to have a bias of $b\sim1.3$ relative to the underlying
dark matter (for $\sigma_8=0.9$), which implies a bias of $b\sim1.2 \pm0.3$
for quasars.  Following Sheth \& Tormen 1999, this implies a minimum
dark matter halo mass of $M_{min}\sim7 \times10^{11} M_\sun$,
corresponding to a mean halo mass (containing one galaxy per halo) of
$M_{200}\sim 3 \times10^{12} M_\sun$ for a concordance cosmology with
$h=0.7$, where $M_{200}$ is the mass within the radius where the
overdensity is 200$\times$ the background density.  Our results here
imply that quasars reside in halos of similar masses as the DEEP2
galaxies.

\subsection{Comparison to Other Observations}

These results are consistent with other findings that AGN cluster
similarly to galaxies \citep{Wake04, Adelberger05a, Constantin06}, but
here are extrapolated to quasar luminosities.  The quasar clustering
amplitude and bias we find here are on the low end of what was
measured in 2dF data using the quasar auto-correlation function
\citep{Porciani04,Croom05}, though within the 3 $\sigma$ errors.
\cite{Porciani04} report that $r_0=4.7 \pm0.7$ for a slope of
$\gamma=1.8$, corresponding to a quasar bias of $1.8 (+0.2/-0.24)$ at
$z=1.06$.  Our clustering amplitude and inferred bias are 1.7$\sigma$ 
and 1.6$\sigma$ lower than these results.  They quote a minimum dark
matter halo mass of $M_{min}=1 \times10^{12} M_\sun$ and a
characteristic mass of $M=1 \times10^{13} M_\sun$, similar to what 
we find within the errors.

\cite{Croom05} measure a redshift-space $\xi(s)$, not a real-space
$\xi(r)$, which is not easy to compare with our results as $\xi(s)$ is
not a power-law and the results therefore depend on the range of
scales fit.  They attempt to model redshift space distortions and
recover \xir, but systematic uncertainties in this modeling make a
direct comparison here difficult.  Their inferred bias, averaged over
the redshift range $0.3<z<2.2$, is $2.02 \pm0.07$, 
2.7$\sigma$ higher than what we measure at $z\sim1$.

\cite{Myers06} employ a lightly different technique in interpreting the 
angular projected correlation function for quasars as a function of 
redshift, where they quote the inferred correlation length at $z=0$.
From their Table 1, using the weighted mean of $r_0(z=0)$ in the
$z_{phot}=0.75$ and $z_{phot}=1.20$ bins (using the $dN/dz$ from
spectra for the redshift distribution of quasars) and using their 
model for the redshift evolution of quasar clustering, the inferred 
 $r_0(z=0.9)=4.23 \pm0.48$ \mpch, 1.5$\sigma$ higher than what is found here.

Overall we find a quasar clustering amplitude and bias that are
$\sim1-2 \sigma$ lower than previous results.  Our measurements 
have comparable error bars to these other studies, which have significantly 
larger quasar samples.  This shows the power of using cross-correlations with
galaxy samples instead of auto-correlations of quasar samples alone.

\subsection{Comparison to Theoretical Models}

Our results can also be compared to theoretical models of how quasars
that are fueled by galaxy mergers should cluster at
$z=1$. \cite{Kauffmann02} model the quasar-galaxy cross-correlation
function using a semi-analytic model in which quasars
are fueled by major galaxy mergers \citep{Kauffmann00}.  
In this model, the peak quasar
luminosity depends on the mass of gas accreted by the black hole, and
the quasar luminosity declines exponentially after the merger event.
This leads to a natural prediction that brighter quasars reside in
more massive galaxies, so that the quasar
clustering amplitude should be luminosity-dependent in a manner which is sensitive to quasar lifetimes.  They also predict that the relative bias of quasars to galaxies should be roughly
scale-independent on scales $r>1$ \mpch, but rises on smaller scales
due to merging events. Their model indicates that the relative bias of
quasars to galaxies decreases at higher redshift and is $\sim1$ at
$z=1$, in accord with the results presented here.  Recent measurements 
of the quasar auto-correlation function do not show a strong dependence of
clustering on luminosity \citep{Croom05}, which may be problematic for
this paradigm, although the errors on the observations are still large
enough ($\sim$30\% in the 2dF data) to be consistent with its predictions.

More recently, \cite{Hopkins05a} present an alternative model for
quasar lifetimes in which bright and faint quasars are in similar physical
systems but are in different stages of their life cycles.  
This work is inspired by numerical simulations of
galaxy mergers \cite{Springel05} which incorporate black hole growth
and feedback.  Whereas the \cite{Kauffmann00} model assumes an
exponential decline of the quasar luminosity with time, in the
\cite{Hopkins05a} scenario quasars spend more time on the
lower-luminosity end of their light curves.  The Hopkins et al. approach is
also able to explain both the optical and X-ray quasar luminosity
functions \citep{Hopkins05b}.

\cite{Lidz06} build on this model, postulating from simulations that
halo mass is strongly correlated with the peak quasar luminosity, but
only indirectly connected to the highly variable instantaneous
luminosity.  This leads to predictions that faint and bright quasars
reside in similar--mass dark matter halos and that quasar clustering
should depend only weakly on luminosity.  Based on this hypothesis,
they estimate the mass distribution of dark matter halos that host
active quasars at $z=2$ and the characteristic dark matter halo masses
of active quasars at $0<z<3$ from the observed quasar luminosity
function.  Their model interprets the observed luminosity
evolution of quasars as a slow 'downsizing' in the quasar population
from $z=2$ to $z=0$, in that at lower redshifts less massive halos
host active quasars.
At $z=1$, the characteristic parent halo mass (defined as
the peak of the lognormal distribution) is estimated to be $M\sim1
\times10^{13} M_\sun$, broadly consistent with our results though
somewhat higher than what we find here.  However, from the inferred
halo mass distribution of quasars they predict that the mean quasar
bias at $z=1$ will be $b=1.95$ (for $\sigma_8=0.9$), significantly different
from our measurement.

In contrast, the semi-analytic model of \cite{Croton06} combines the 
\cite{Kauffmann00} prescription of merger-driven black hole growth 
with an independent low energy `radio mode' heating mechanism efficient at 
late times.  Because this model assumes that some fraction of cold gas 
must be present to trigger a `quasar' phase during a galaxy merger, 
gas-free galaxies are not expected to shine as quasars.   Thus, this
model predicts that active quasars should only occur below a maximum dark 
matter halo mass where the heating mechanism has yet to switch on, 
the so-called 'quenching mass'.  Above the quenching mass, gas in the 
IGM will be hot enough to 
suppress the infall of new cold gas onto galaxies, so that black holes 
are only fed slowly  \citep[e.g.,][]{Churazov05}. 
Common, group-scale halos will generally pass this mass threshold 
($\sim 3 \times 10^{11} M_\sun$) at $z \sim1-2$. At lower redshifts, then,
only those mergers which occur outside of group environments will contain the 
requisite cold gas to fuel a strong AGN. Simply put, in this scenario quasar 
activity is quenched in dense environments for much the same reasons that 
star formation is, and our finding that quasars at $z\sim1$ cluster very 
similarly to bright star-forming galaxies at the same redshifts is not 
necessarily surprising.

\subsection{Implications for Galaxy Evolution}

Our findings support a picture where quasars at $z\sim1$ are a transient phase in the life of normal galaxies, given that they cluster similarly.
This does not necessarily imply that {\bf all} galaxies have hosted a
quasar at some point, however, but just that the galaxies that do at
 $z\sim 1$ are found in fairly typical dark matter halos.  
\cite{Porciani04} compare the 
observed abundance of quasars to the number density of dark matter halos with masses corresponding to  the observed clustering properties;  they conclude that at $z=1$ only 1\% of all 
potential host dark matter halos actually harbor a quasar.
A direct comparison of the quasar number density derived from the 
2dF QSO luminosity function \citep{Boyle00} at $z=0.8$ to the number density of
DEEP2 galaxies \citep{Willmer06} leads to the conclusion that roughly
one out of every $\sim400$ DEEP2 galaxies hosts a quasar with $M_B<-22$,
and twice as many host a quasar with $M_B<-21$.

If quasars are fueled by galaxy mergers, then the clustering amplitude
on small scales ($\lesssim0.1$ \mpch) should rise relative to large
scales.  There is no significant scale-dependence seen for the scales
investigated here ($0.1 < r_p < 10$ \mpch), but \cite{Hennawi05} find
that on scales $r_p<100 h^{-1} kpc$ there is a strong increase in
the quasar clustering amplitude above a power-law extrapolation 
in the range $0.7 < z < 3.0$.  
Similarly, a small scale excess of L$^\ast$ galaxies was detected around
low redshift ($z < 0.4$) AGN by \cite{Serber06}. Our
quasar sample is too small to provide clustering measures with
reasonable errors on these scales and so we are not able to address
this question here (there are only a total of 5 galaxies
contributing to the smallest $r_p$ bin shown).  However, the quasar-galaxy
cross-correlation function is much better suited to address such 
small-scale behavior than the quasar auto-correlation function, as
the galaxy population has a much higher number density than the quasar
population.  Much improved statistics could be obtained by surveying
the galaxy population densely in the neighborhood of known quasars.

Models that propose that major mergers between blue galaxies fuel
quasars, which in turn quench star-formation and lead to the formation
of red-sequence galaxies, have to match the observed clustering of
quasars relative to galaxies.  We find evidence that quasars cluster
more like blue galaxies than red, which may pose a problem if most red
galaxies form rapidly from quasars.  If quasars are in blue galaxies
that will soon migrate to the red sequence, then we might expect them
to have an intermediate clustering amplitude between the average blue
and red galaxy populations, which is consistent with, but not favored
by, our findings.  One could better reconcile the data with these
models if different subsets of the red galaxy population have
different clustering; e.g., older red galaxies may be more clustered
than galaxies that have recently undergone a quasar phase and joined
the red sequence.  We note, however, that both the DEEP2 and COMBO-17
surveys find that the red sequence population grows rapidly from
$z\sim 1$ to $z\sim 0$ \citep{Faber06}.  This implies that the red
sequence population is dominated by number by relatively young
galaxies at $z\sim 1$; this is also found from stellar population
measurements by \cite{Schiavon06}.  This then requires that the
rare, older red sequence galaxies must be that much more strongly
clustered than younger ones to match their observed overall
correlation function.

In this paper we have used a relatively small quasar sample (52 objects)
to show that quasars have comparable clustering properties and 
reside in similar environments and dark matter halos as DEEP2 galaxies.  
We find tentative ($2\sigma$) evidence that the quasar clustering amplitude 
matches that of blue galaxies at $z\sim1$ more than red galaxies, and 
that quasars are only modestly biased relative to dark matter  
($b\sim1.3$).  
This paper shows the potential of cross-correlation 
techniques and points the way to future studies.  
We hope to improve on these results using additional 
quasars that are spectroscopically confirmed in the DEEP2 fields, particularly
radio and X-ray sources identified in the Extended Groth Strip, a 
DEEP2 field with considerable multi-wavelength data.  With smaller 
measurement errors we may be able to distinguish between different models 
of quasar formation and evolution, as discussed above.

\acknowledgements We would like to thank Daniel Eisenstein and Michael
Strauss for useful conversations and Darren Croton, Sandy Faber, 
Ben Weiner and Christopher Willmer for comments on earlier drafts.  
We also thank Scott Burles, Gordon Richards and Todd Boroson for 
helpful discussions about the accuracy of quasar redshifts and the
anonymous referee for a helpful report.  This project
was supported by the NSF grant AST-0071048. ALC, JFH and JAN are
supported by NASA through Hubble Fellowship grants HF-01182.01-A,
HF-0117.01-A and HST-HF-011065.01-A, respectively, awarded by the
Space Telescope Science Institute, which is operated by the
Association of Universities for Research in Astronomy, Inc., for NASA,
under contract NAS 5-26555.

The DEIMOS spectrograph was funded by a grant from CARA (Keck
Observatory), an NSF Facilities and Infrastructure grant (AST92-2540),
the Center for Particle Astrophysics and by gifts from Sun
Microsystems and the Quantum Corporation.  The DEEP2 Redshift Survey
has been made possible through the dedicated efforts of the DEIMOS
staff at UC Santa Cruz who built the instrument and the Keck
Observatory staff who have supported it on the telescope.  The data
presented herein were obtained at the W.M. Keck Observatory, which is
operated as a scientific partnership among the California Institute of
Technology, the University of California and the National Aeronautics
and Space Administration. The Observatory was made possible by the
generous financial support of the W.M. Keck Foundation. The DEEP2 team
and Keck Observatory acknowledge the very significant cultural role
and reverence that the summit of Mauna Kea has always had within the
indigenous Hawaiian community and appreciate the opportunity to
conduct observations from this mountain.

Funding for the Sloan Digital Sky Survey (SDSS) has been provided by
the Alfred P. Sloan Foundation, the Participating Institutions, the
National Aeronautics and Space Administration, the National Science
Foundation, the U.S. Department of Energy, the Japanese
Monbukagakusho, and the Max Planck Society. The SDSS Web site is
http://www.sdss.org/.  The SDSS is managed by the Astrophysical
Research Consortium (ARC) for the Participating Institutions. The
Participating Institutions are The University of Chicago, Fermilab,
the Institute for Advanced Study, the Japan Participation Group, The
Johns Hopkins University, the Korean Scientist Group, Los Alamos
National Laboratory, the Max-Planck-Institute for Astronomy (MPIA),
the Max-Planck-Institute for Astrophysics (MPA), New Mexico State
University, University of Pittsburgh, University of Portsmouth,
Princeton University, the United States Naval Observatory, and the
University of Washington.

%\bibliography{references}

\begin{deluxetable}{lcccccccccccc}
\tablecolumns{10}
\tablewidth{0pc}
\tablecaption{SDSS Quasars in the DEEP2 Fields
\label{table:qsos}}
\tablehead{Name & z & $M_{B}$ & RA (J2000) & Dec (J2000) & $u$ &$g$ & $r$ & $i$ & $z$}
\startdata
SDSSJ0226$+$0022 &   1.008  &   \phn-23.0  &   02:26:55.25 &   $+$00:22:11.2 &   20.44  &   20.32  &   20.11  &   20.17  &   20.14 \\
SDSSJ0226$+$0023 &   0.984  &   \phn-22.9  &   02:26:56.52 &   $+$00:23:47.7 &   20.48  &   20.38  &   20.22  &   20.24  &   20.08 \\
SDSSJ0227$+$0048 &   1.108  &   \phn-23.2  &   02:27:26.10 &   $+$00:48:27.6 &   20.33  &   20.44  &   20.17  &   20.21  &   20.28 \\
%SDSSJ0227$+$0047 &   1.48  &   \phn-23.9  &   02:27:30.14 &   $+$00:47:33.7 &   20.89  &   20.74  &   20.23  &   19.99  &   19.96 \\
SDSSJ0228$+$0030 &   1.013  &   \phn-23.2  &   02:28:37.60 &   $+$00:30:10.3 &   20.10  &   19.99  &   19.79  &   20.01  &   19.90 \\
SDSSJ0228$+$0033 &   0.768  &   \phn-22.8  &   02:28:38.62 &   $+$00:33:20.2 &   20.04  &   19.74  &   19.73  &   19.86  &   19.62 \\
SDSSJ0228$+$0046 &   1.287  &   \phn-25.2  &   02:28:39.33 &   $+$00:46:23.0 &   19.25  &   18.88  &   18.64  &   18.52  &   18.50 \\
SDSSJ0228$+$0030 &   0.720  &   \phn-24.3  &   02:28:41.08 &   $+$00:30:49.5 &   18.24  &   18.01  &   18.10  &   18.21  &   18.05 \\
SDSSJ0229$+$0039 &   1.209  &   \phn-23.7  &   02:29:08.58 &   $+$00:39:08.1 &   20.85  &   20.52  &   20.00  &   19.85  &   20.10 \\
%SDSSJ0229$+$0027 &   1.49  &   \phn-24.3  &   02:29:38.18 &   $+$00:27:16.7 &   20.12  &   20.09  &   19.90  &   19.68  &   19.69 \\
SDSSJ0229$+$0046\tablenotemark{a} &   0.787  &   \phn-22.5  &   02:29:59.59 &   $+$00:46:32.0 &   21.10  &   20.73  &   20.54  &   20.20  &   19.89 \\   % object #6 in DEEP2
SDSSJ0230$+$0049 &   1.284  &   \phn-24.5  &   02:30:56.75 &   $+$00:49:33.6 &   19.33  &   19.36  &   19.16  &   19.16  &   19.25 \\
SDSSJ0231$+$0024 &   1.049  &   \phn-22.9  &   02:31:01.49 &   $+$00:24:03.1 &   20.74  &   20.51  &   20.22  &   20.38  &   20.74 \\
SDSSJ0231$+$0051\tablenotemark{a} &   1.211  &   \phn-24.0  &   02:31:15.42 &   $+$00:51:42.2 &   19.95  &   19.91  &   19.63  &   19.54  &   19.38 \\   % object #1 in DEEP2
SDSSJ0231$+$0044 &   1.266  &   \phn-24.1  &   02:31:23.80 &   $+$00:44:25.9 &   20.41  &   20.19  &   19.67  &   19.49  &   19.46 \\
%SDSSJ1413$+$5212 &   1.21  &   \phn-24.5  &   14:13:41.13 &   $+$52:12:20.4 &   19.24  &   19.10  &   18.93  &   18.99  &   19.07 \\
SDSSJ1415$+$5205\tablenotemark{a} &   0.985  &   \phn-24.5  &   14:15:33.90 &   $+$52:05:58.2 &   18.97  &   18.84  &   18.63  &   18.71  &   18.66 \\
SDSSJ1416$+$5218 &   1.284  &   \phn-25.9  &   14:16:42.44 &   $+$52:18:12.9 &   18.04  &   18.04  &   17.82  &   17.78  &   17.88 \\
%SDSSJ1418$+$5223 &   1.12  &   \phn-24.7  &   14:18:38.36 &   $+$52:23:59.5 &   18.95  &   18.83  &   18.63  &   18.67  &   18.75 \\
%SDSSJ1420$+$5313 &   0.74  &   \phn-23.2  &   14:20:56.84 &   $+$53:13:07.6 &   21.06  &   20.75  &   20.22  &   19.43  &   19.12 \\
SDSSJ1646$+$3503 &   0.859  &   \phn-23.4  &   16:46:34.71 &   $+$35:03:17.6 &   19.98  &   19.47  &   19.41  &   19.51  &   19.41 \\
SDSSJ1647$+$3505 &   0.861  &   \phn-22.8  &   16:47:33.24 &   $+$35:05:41.7 &   20.20  &   19.95  &   19.92  &   20.14  &   20.10 \\
SDSSJ1649$+$3452 &   0.739  &   \phn-24.2  &   16:49:12.23 &   $+$34:52:52.6 &   19.16  &   18.68  &   18.51  &   18.42  &   18.23 \\
SDSSJ1650$+$3451 &   1.300  &   \phn-24.5  &   16:50:46.31 &   $+$34:51:38.4 &   19.49  &   19.50  &   19.23  &   19.23  &   19.29 \\
SDSSJ1651$+$3506\tablenotemark{a} &   0.753  &   \phn-22.5  &   16:51:16.41 &   $+$35:06:35.3 &   20.28  &   20.01  &   20.02  &   20.10  &   19.83 \\ % DEEP2: & 20.32 & 20.20 & 19.88 \\ 
SDSSJ1652$+$3500 &   1.381  &   \phn-23.9  &   16:52:49.27 &   $+$35:00:57.0 &   20.29  &   20.13  &   19.96  &   19.94  &   20.01 \\
SDSSJ2325$+$0019 &   1.203  &   \phn-24.2  &   23:25:36.23 &   $+$00:19:08.6 &   19.66  &   19.67  &   19.25  &   19.33  &   19.44 \\
%SDSSJ2325$-$0005 &   1.41  &   \phn-25.3  &   23:25:56.96 &   $-$00:05:00.0 &   18.80  &   18.77  &   18.60  &   18.58  &   18.67 \\
SDSSJ2326$+$0009 &   1.034  &   \phn-23.3  &   23:26:26.15 &   $+$00:09:22.2 &   19.25  &   19.94  &   19.82  &   19.94  &   20.13 \\
SDSSJ2326$-$0003 &   1.277  &   \phn-23.7  &   23:26:32.90 &   $-$00:03:26.7 &   20.06  &   20.19  &   19.88  &   19.99  &   20.11 \\
SDSSJ2326$+$0021 &   1.258  &   \phn-23.4  &   23:26:34.71 &   $+$00:21:49.7 &   20.55  &   20.58  &   20.26  &   20.19  &   20.33 \\
SDSSJ2326$-$0005 &   1.030  &   \phn-22.4  &   23:26:38.12 &   $-$00:05:24.7 &   21.48  &   21.18  &   20.74  &   20.84  &   20.24 \\
SDSSJ2327$-$0002 &   1.235  &   \phn-23.8  &   23:27:23.69 &   $-$00:02:43.2 &   20.14  &   20.07  &   19.77  &   19.82  &   19.83 \\
%SDSSJ2327$+$0022 &   1.49  &   \phn-25.9  &   23:27:34.74 &   $+$00:22:34.0 &   18.38  &   18.41  &   18.16  &   18.06  &   18.10 \\
SDSSJ2327$+$0006 &   0.884  &   \phn-23.6  &   23:27:42.67 &   $+$00:06:53.9 &   18.76  &   19.22  &   19.19  &   19.32  &   19.31 \\
SDSSJ2327$-$0000 &   0.986  &   \phn-22.7  &   23:27:57.24 &   $-$00:00:35.9 &   20.82  &   20.86  &   20.36  &   20.46  &   20.30 \\
%SDSSJ2328$+$0022 &   1.31  &   \phn-26.0  &   23:28:20.38 &   $+$00:22:38.2 &   17.93  &   17.89  &   17.69  &   17.72  &   17.77 \\
SDSSJ2329$+$0012 &   1.210  &   \phn-24.6  &   23:29:03.41 &   $+$00:12:26.9 &   19.99  &   19.48  &   19.08  &   18.89  &   18.97 \\
SDSSJ2329$+$0009 &   0.881  &   \phn-23.1  &   23:29:51.46 &   $+$00:09:42.8 &   19.53  &   20.01  &   19.84  &   19.88  &   19.77 \\
SDSSJ2330$+$0017 &   0.705  &   \phn-23.6  &   23:30:20.72 &   $+$00:17:27.6 &   19.57  &   18.96  &   18.93  &   18.82  &   18.80 \\
SDSSJ2330$+$0008 &   0.994  &   \phn-23.8  &   23:30:23.48 &   $+$00:08:11.9 &   18.92  &   19.47  &   19.28  &   19.34  &   19.38 \\
%SDSSJ2332$+$0019 &   1.47  &   \phn-23.5  &   23:32:14.84 &   $+$00:19:30.2 &   20.92  &   20.85  &   20.75  &   20.42  &   20.26 \\
SDSSJ2332$+$0001\tablenotemark{a} &   0.716  &   \phn-22.1  &   23:32:30.42 &   $+$00:01:37.7 &   20.01  &   20.34  &   20.29  &   20.35  &   19.86 \\
SDSSJ2333$-$0004\tablenotemark{a} &   0.697  &   \phn-22.6  &   23:33:15.90 &   $-$00:04:52.9 &   20.11  &   19.79  &   19.77  &   19.85  &   19.66 \\
SDSSJ2333$-$0003 &   0.919  &   \phn-22.4  &   23:33:29.01 &   $-$00:03:08.2 &   21.20  &   20.96  &   20.81  &   20.61  &   20.49 \\
\enddata
\tablenotetext{a}{Also observed by the DEEP2 Galaxy Redshift Survey}
\tablecomments{ \footnotesize 
  The redshift and B-band absolute magnitude of the quasar 
  are designated by z and $M_{\rm B}$, respectively. 
  Extinction corrected SDSS five band PSF photometry are
  given in the columns $u$, $g$, $r$, $i$, and $z$. Absolute magnitudes
  $M_{\rm B}$ are computed from the cross filter K-correction $K_{Bi}(z)$,
  between apparent magnitude $i$ and absolute magnitude $B$, which 
  was computed from the SDSS composite quasar spectrum of \citet{Vanden01}.  
}
\end{deluxetable}

\clearpage
\thispagestyle{empty}

\begin{deluxetable}{rccccccccccccccc}
\tablecolumns{13}
\tablewidth{0pc}
\tablecaption{Additional DEEP2 quasars and broad-lined AGN
\label{table:agn}}
\tablehead{Name & z & $M_{B}$ & RA (J2000) & Dec (J2000) & $u$ &$g$ & $r$ & $i$ & $z$ & $B$ & $R$ & $I$ }
\startdata
DEEP2J0226$+$0033 &   0.764  &   \phn-21.1  &   02:26:50.11 &   $+$00:33:04.5 &   22.78  &   22.71  &   22.40  &   21.49  &   20.93  &   22.40  &   21.49  &   20.87 \\
DEEP2J0228$+$0034 &   0.708  &   \phn-21.4  &   02:28:13.57 &   $+$00:34:55.5 &   21.78  &   21.19  &   21.16  &   21.05  &   20.76  &   21.07  &   21.06  &   20.63 \\
DEEP2J1421$+$5306 &   1.327  &   \phn-23.8  &   14:21:16.68 &   $+$53:06:07.4 &   22.51  &   21.23  &   20.30  &   19.92  &   19.77  &   21.30  &   20.05  &   19.61 \\
DEEP2J1649$+$3508 &   0.769  &   \phn-20.4  &   16:49:08.03 &   $+$35:08:08.6 &   22.57  &   21.69  &   21.51  &   21.64  &   21.70  &   22.29  &   22.23  &   21.49 \\
DEEP2J1651$+$3443 &   0.731  &   \phn-21.7  &   16:51:04.66 &   $+$34:43:06.3 &   21.45  &   21.26  &   21.12  &   20.82  &   20.66  &   21.19  &   20.87  &   20.55 \\
%DEEP2J1651$+$3506 &   0.753  &   \phn-22.5  &   16:51:16.41 &   $+$35:06:35.3 &   20.28  &   20.01  &   20.02  &   20.10  &   19.83  &   20.32  &   20.20  &   19.88 \\
% listed in SDSS table as well - i accidentally doubled-up on this one
DEEP2J1652$+$3447 &   0.804  &   \phn-21.3  &   16:52:36.39 &   $+$34:47:22.4 &   21.83  &   21.41  &   21.20  &   21.12  &   20.92  &   21.48  &   21.37  &   21.07 \\
DEEP2J1652$+$3506 &   0.841  &   \phn-20.8  &   16:52:56.13 &   $+$35:06:32.3 &    ...   &   ...    &   ...    &   ...    &    ...   &   23.54  &   21.95  &   21.16 \\
DEEP2J2327$+$0017 &   0.954  &   \phn-22.8  &   23:27:07.68 &   $+$00:17:24.7 &   21.41  &   20.98  &   20.54  &   20.27  &   19.96  &   21.17  &   20.40  &   19.89 \\
DEEP2J2327$-$0001 &   0.771  &   \phn-21.7  &   23:27:40.09 &   $-$00:01:44.0 &   21.46  &   21.16  &   20.98  &   20.96  &   20.69  &   21.19  &   21.09  &   20.67 \\
DEEP2J2328$+$0012 &   0.748  &   \phn-22.0  &   23:28:17.64 &   $+$00:12:07.4 &   20.79  &   20.66  &   20.57  &   20.56  &   19.92  &   21.25  &   20.75  &   20.28 \\
DEEP2J2329$+$0018 &   1.387  &   \phn-21.1  &   23:29:04.80 &   $+$00:18:08.6 &   22.00  &   22.17  &   22.22  &   22.07  &   21.46  &   22.91  &   22.73  &   22.61 \\
DEEP2J2329$+$0006 &   0.743  &   \phn-21.5  &   23:29:22.77 &   $+$00:06:22.2 &   21.75  &   21.39  &   21.25  &   21.10  &   21.23  &   21.09  &   21.05  &   20.87 \\
DEEP2J2329$+$0015 &   1.391  &   \phn-22.2  &   23:29:29.52 &   $+$00:15:49.0 &   22.63  &   22.65  &   22.44  &   22.57  &   22.30  &   21.76  &   21.68  &   21.74 \\
DEEP2J2329$+$0025 &   1.387  &   \phn-20.9  &   23:29:51.92 &   $+$00:25:18.0 &   22.84  &   23.01  &   23.07  &   22.75  &   21.66  &   23.09  &   22.97  &   22.88 \\
DEEP2J2330$+$0014 &   1.387  &   \phn-21.0  &   23:30:06.28 &   $+$00:14:59.4 &     ...  &   ...    &   ...    &   ...    &   ...    &   23.63  &   22.80  &   22.19 \\
DEEP2J2333$+$0005 &   1.386  &   \phn-23.5  &   23:33:55.08 &   $+$00:05:46.3 &   20.38  &   20.95  &   20.32  &   20.37  &   20.42  &   20.44  &   20.37  &   20.45 \\
\enddata
%\tablecomments{}
\end{deluxetable}

\clearpage
\setlength{\hoffset}{-15mm}

\begin{deluxetable}{lcccccc}
\tablewidth{0pt}
\tablecaption{Power-law fits to Clustering Results for Quasar Samples}
\tablehead{
\colhead{}      & \multicolumn{3}{c}{($0.1<r_p<10$ \mpch)} & \multicolumn{3}{c}{($1<r_p<10$ \mpch)} \\
\colhead{Sample}&\colhead{$r_0$}&\colhead{$\gamma$} &\colhead{Relative}&\colhead{$r_0$}&\colhead{$\gamma$} &\colhead{Relative}\\
\colhead{}      &\colhead{\mpch}&\colhead{}         &\colhead{Bias}    &\colhead{\mpch}&\colhead{}         &\colhead{Bias} \\
}
\startdata
SDSS $M_B<-22$        & $2.95 \pm0.44$ & $1.72 \pm0.26$ & $0.80 \pm0.35$ & $3.20 \pm0.51$ & $1.86 \pm0.24$ & $0.75 \pm0.28$ \\
SDSS+DEEP2 $M_B<-22$  & $3.35 \pm0.50$ & $1.83 \pm0.27$ & $0.90 \pm0.26$ & $3.56 \pm0.57$ & $1.96 \pm0.25$ & $0.85 \pm0.27$ \\
SDSS+DEEP2 All Quasars  & $3.31 \pm0.70$ & $1.55 \pm0.23$ & $0.86 \pm0.23$ & $3.39 \pm0.71$ & $1.53 \pm0.21$ & $0.89 \pm0.24$ 
\enddata
\label{restable}
\end{deluxetable}

\clearpage
\setlength{\hoffset}{0mm}

\begin{deluxetable}{lrc}
\tablewidth{0pt}
\tablecaption{Environment Measures for Quasars and DEEP2 Galaxies}
\tablehead{
\colhead{Sample} & \colhead{No. Objects} & \colhead{log(1+$\delta$)} \\
}
\startdata
All \ DEEP2 Galaxies   & 16,761 & $0.031 \pm0.005$ \\
Blue DEEP2 Galaxies  & 14,282 & $0.004 \pm0.005$ \\
Red \ DEEP2 Galaxies   & 2,479  & $0.190 \pm0.013$ \\
SDSS+DEEP2 Quasars & 52     & $0.010 \pm0.089$\\
\enddata
\label{envtable}
\end{deluxetable}

\begin{figure}
\centerline{\scalebox{0.6}{\includegraphics{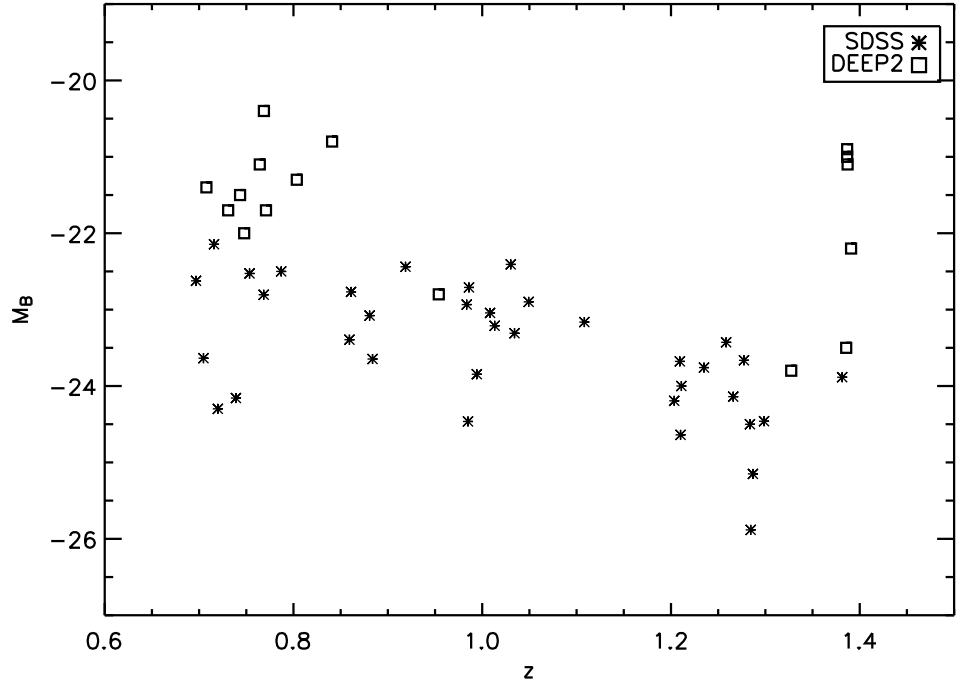}}}
\caption{The redshifts and absolute $M_B$ magnitudes for broad-lined 
quasars in our sample, from both the SDSS (astericks) and DEEP2 
(squares) surveys.  We define three quasar samples for cross-correlating with 
the DEEP2 galaxies, as detailed in the text.  The lack of quasars identified
in DEEP2 between $0.9<z<1.3$ is due to the limited spectral range of the
high-resolution DEEP2 spectra; there are no strong quasar emission lines
observed at these redshifts.
\label{QSOprops}}
\end{figure}

\begin{figure}
\centerline{\scalebox{0.6}{\includegraphics{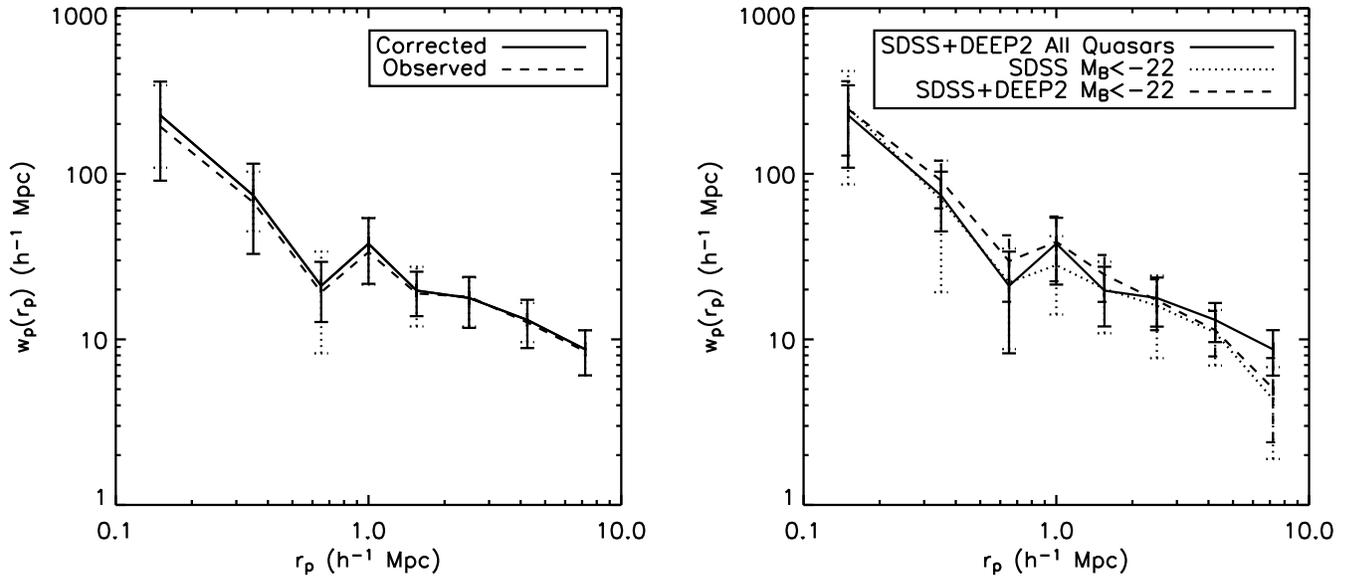}}}
\caption{
Left: The projected cross-correlation function, \wprp, between
SDSS quasars and the DEEP2 galaxy sample. 
The observed correlation function is shown as 
a dashed line, while the solid line shows results corrected for the DEEP2 
target slitmask algorithm.   The solid error bars are 
derived from jacknife resampling, while the dotted error bars reflect the 
standard deviation in the mock galaxy catalogs.   
Right: The projected correlation function for all three quasar samples,
shown with jacknife errors.
\label{wprp}}
\end{figure}

\begin{figure}
\centerline{\scalebox{0.6}{\includegraphics{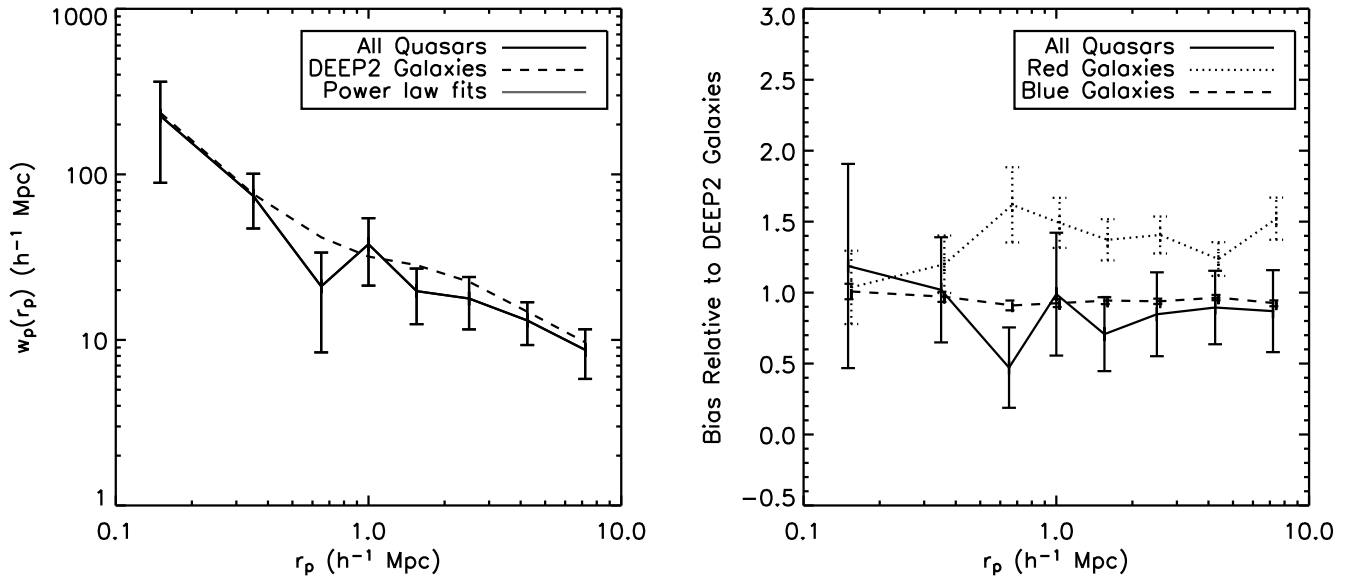}}}
\caption{Left: The projected cross-correlation function between SDSS
  and DEEP2 quasars and all DEEP2 galaxies is shown as a solid line, while 
the dashed line shows the auto-correlation function of DEEP2 galaxies within $\Delta
  z=0.1$ of the quasars (see text for details). 
Right: The solid line shows the relative bias between quasars 
and all DEEP2 galaxies as a function of scale, while the dotted (dashed) 
lines shows the relative bias between red (blue) galaxies and all 
galaxies in the DEEP2 data. 
\label{bias}}
\end{figure}

\end{document}